\documentclass[12pt]{article}
\usepackage{amssymb}
\usepackage{amsfonts}
\usepackage{amsmath,amsthm}
\usepackage{graphics,epsfig}
\usepackage{subfigure}
\usepackage{graphicx,epsfig, color }
\oddsidemargin 10mm \numberwithin{equation}{section}
\def\be{\begin{equation}}
\def\ee{\end{equation}}
\begin{document}
\begin{center}
{{\bf {  Physics-informed deep learning for three dimensional
black holes
 \\
}} \vskip 1 cm
 {  E. Yaraie $^{a,b}$
        \footnote{E-mail address: eyaraie@semnan.ac.ir}} H. Ghaffarnejad $^{a}$\footnote{E-mail address:
hghafarnejad@semnan.ac.ir}  and M. Farsam $^{a,b}$
        \footnote{E-mail address: mhdfarsam@semnan.ac.ir}}\\
   \vskip 0.5 cm
   \textit{$^a$ Faculty of Physics, Semnan University, P.C. 35131-19111, Semnan,
   Iran}\\
   \textit{$^b$Instituut-Lorentz for Theoretical Physics, ITP, Leiden University, Niels Bohrweg 2, Leiden 2333 CA, The
   Netherlands}\\
    \end{center}
\begin{abstract}
According to AdS/DL (Anti de Sitter/ Deep Learning) correspondence
given by \cite{Has}, in this paper with a data-driven approach and
leveraging holography principle we have designed an artificial
neural network architecture to produce metric field of planar BTZ
and quintessence black holes. Data has been collected by choosing
minimally coupled massive scalar field with quantum fluctuations
and try to process two emergent and ground-truth metrics versus
the holographic parameter which plays role of depth of the neural
network. Loss or error function which shows rate of deviation of
these two metrics in presence of penalty regularization term
reaches to its minimum value when values of the learning rate
approach to the observed steepest gradient point. Values of the
regularization or penalty term of the quantum scalar field has
critical role to matching this two mentioned metric.
 Also we design an algorithm which helps us to
find optimum value for learning parameter and at last we
understand that loss function convergence heavily depends on the
number of epochs and learning rate.
\end{abstract}
\section{Introduction}
After prediction of black holes evaporation in presence of quantum
matter field by Hawking \cite{Ha} and the black hole entropy by
Bekenestein \cite{bec1} which is related to surface gravity of the
black hole, Susskind and t'Hooft stated that the theory of quantum
gravity within any region is encoded on the surface of that region
\cite{Ho,Suss} which is called now as the holographic principle.
The best successful theory so far for the holographic principle is
the Anti de Sitter-conformal field theory (AdS/CFT) correspondence
which is proposed by Maldacena \cite{Mal} for the first time. This
correspondence has two consequences such that the quantum gravity
in each slice of AdS spacetime is explained by the data on the
boundary slice and information which lives on the boundary evolves
between the slices of the AdS spacetime by the Hamiltonian of
conformal invariant quantum fields. The study of AdS black holes
in $d<4$ dimensions are conducted in a variety of ways \cite{btz}
\cite{kleber}. For instance one can see \cite{son,abdal} for
understanding of the dual field theory in the context of AdS/CFT
correspondence and \cite{Kiselev,ours1,ours2,ours3,Chen,olov} for
studying of effects of the quintessence fields in 1+2 dimensional
black holes spacetimes. Compared to 4D case, BTZ black hole has
certain good theoretical properties, e.g. this is just an example,
not proposal: turning on angular momentum is simpler compared to
4D case, and something new may be checked by utilizing it.
\\
Deep neural network which is known as deep structured learning is
part of a broader family of machine learning methods based on
artificial neural networks with representation learning. This
 is shown that is extended to be applicable for more
branches of physical science such as the gravitational and the
cosmological context (see \cite{Car} for a good review). For
instance one can see some published works about application of
deep learning method related with gravity as follows: Yong Yang et
al used deep learning method to determine atmospheric parameters
of white dwarf stars recently \cite{YY}. Christopher J. Shallue
and Andrew Vanderburg also used deep learning method to identify
exoplanets \cite{Chri}. Matsuoka et al, apply the deep learning
method to estimate parameters of atmospheric gravity wave in
reanalysis data sets \cite{Matsu}. In fact, neural networks that
work according to the laws of physics are called physics-informed
neural networks (see \cite{Geo} and references therein). This kind
of learning algorithm is inspired by information processing and it
is distributed by communication nodes in biological systems.
 Artificial neural networks models have been
used since the 1950s \cite{rosenblatt} and flourished in the 2000s
\cite{Hinton}.
 It is composed of multiple layers to progressively extract higher-level features from the raw
 input and delivering an output. With respect to the task at hand, the output could be have discrete value or continuous  value
 \cite{Hinton,LeCun}.
  Recent
breakthrough results in computer vision, natural language
processing speech recognition, biomedicine and many other domains
have produced a massive interest in this direction
\cite{vision,bio,speech,nlp}.  Hashimoto et al \cite{Has} in their
work presented recently a deep neural network representation for
the AdS/CFT correspondence. They demonstrated the emergence of the
bulk metric function via the learning process for given data sets
of response in boundary quantum field theories. In this approach
the emergent radial (holographic) direction of the bulk is
identified with the depth of the layers, and the network itself is
interpreted as a bulk geometry. They showed their network provides
a data-driven holographic modeling of strongly coupled systems. By
using $\phi^4$ scalar potential for a minimally coupling dynamical
scalar field with mass parameter $m$ moving on curved spacetimes
with a black hole horizon they demonstrated their deep learning
(DL) framework determine the background metric by fitting given
response data. Their proposal has two steps as follows: At first
step they showed that, from boundary data generated by the AdS
Schwarzschild spacetime, the network can reproduce the metric. At
the second step they demonstrated that the network with
experimental data as an input can determine the bulk metric, the
mass and the quadratic coupling of the holographic model. In the
paper \cite{Has} they studied also the experimental data of
magnetic response of a strongly correlated material Sm0.6Sr0.4MnO3
which has strong quantum fluctuations. At last they showed that
their AdS/DL correspondence not only enables gravity modeling of
strongly correlated systems, but also sheds light on a hidden
mechanism of the emerging space in both AdS and DL. Precedence and
novelty of their work is because that for a quantum system given
we do not know whether its gravity dual exists and how we can
construct a holographic model? In fact for phenomenology, the
holographic modelings were successful only for restricted class of
systems in which symmetries are manifest, mainly because the
mechanism of how the holography works is still unknown. While
conventional holographic modeling starts with a given bulk gravity
metric, Hashimoto et al novel DL method \cite{Has} solves the
inverse problem which means data of a boundary QFT calculates a
suitable bulk metric function by assuming the existence of a black
hole horizon. To do so we should provide a deep neural network
representation of a scalar field equation moving in curved
spacetime. The discretized holographic AdS radial direction is the
deep layers. The weights of the neural network are identified with
metric of the curved spacetime. The input response data is at the
boundary of AdS, and the output binomial data is the black hole
horizon condition. Therefore, a successful machine learning
results in a concrete metric of a holographic modeling of the
system measured by the experiment. This is all which is called as
AdS/DL correspondence of a deep neural network by Hashimoto et al.
When stress tensor of scalar field has zero barotropic index $w=0$
then the 3D black hole reads as planar BTZ black hole while with
non vanishing barotropic index $-1<w\leq-\frac{1}{3}$ the 3D black
hole is called as quintessence black hole which we like to produce
them by using  method of Hashimoto et al.  The paper is organized
as follows:\\ In section 2 we present brief review of
architecturing deep neural network and developing deep neural
learning model. In section 3 we provide a brief review of 1+2
dimensional BTZ black hole metric solution. Then we investigate
correspondence between metric components and parameters of deep
neural network for the BTZ planar black hole such that the black
hole could feed with in input layer by corresponding boundary data
which is labeled with respect to the horizon boundary conditions.
Then when data is propagating towards the black hole horizon, the
spacetime metric is being reproduced. Section 4 is dedicated to
the network architecture, training implementation and data
setting. In the last section we investigate conclusion and outlook
of the work.
\section{Artificial neural network}
\begin{figure}[tbp] \centering
   \includegraphics[scale=1]{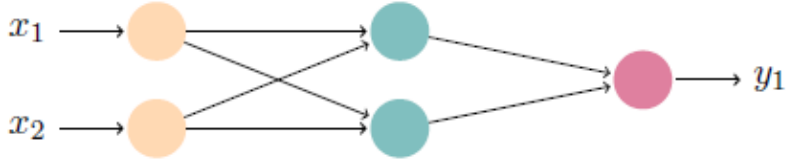}
\caption{\label{z} Schematic diagram of a simple neural network in
which activation function is shown with the colors so that pink,
blue and violet correspond with input, hidden and output neuron
layers respectively. Weights are shown with solid lines. Input
data is $x_{1,2}$ and output one is $y_1.$ }
\end{figure}

A neural network, also sometimes is called an artificial neural
network, is a kind of processing structure which their name and
structure are inspired by the human brain,  mimicking the way
where the biological neurons signal to one another. Basic building
block of a neural network is made in fact by a neuron.  We show
schematic diagram of a simple neural network in figure \ref{z}. In
this figure the artificial neuron takes all the inputs $x_{1,2}$,
weights $W$ (shown with solid lines) which is a linear
transformation
 between vector components of the
neuron as $x_i\to \Sigma_jW_{ij}x_j$, aggregates (not shown) and
an activation function $x_i\to \varphi(x_i)$
 which is usually a nonlinear transformation on the vector components of the neuron $x_i$ such that it should
 deliver the output of the neuron at each layer. In fact the activation function controls
 value of the output when the neuron is activated.
  A row of neurons is called layer and a network can have multiple layers.
   Input layer receives data $x_i$ and delivers output to next layer via two above mentioned transformations as $x_k\to\varphi(W_{kl}x_l)$ and
    final layer is responsible for delivering values which correspond to result demanded
   for the problems such that regression, classification and etc. Layers located between first and last ones are called hidden layers.
In general for N layers a deep feed-forward neural network can be
constructed as follows.
\begin{equation}
y(x^{(1)})=f_i \varphi (W_{ij}^{(N-1)}\varphi
(W_{jk}^{(N-2)}\cdots \varphi (W_{lm}^{(1)}x_m^{(1)}))) \label{ff}
\end{equation}
where $f_i$ means activation function $x_i\to\varphi(x_i)$ but at
last layer which delivers to the target $y(x^{(1)})$. In the
learning process, the variables of the Network
$(f_i,W_{ij}^{(n)})$ for $n=1,2,\cdots N-1$ are updated by a
gradient descent method with a given loss or error function
\begin{equation}
E=\sum_{data}|y(\bar{x}^{(1)})-\bar{y}|+E_{reg}(W).\label{loss11}
\end{equation}
Here the sum is over the whole set of pairs
$\{(\bar{x}^{(1)},\bar{y})\}$ of the input data $\bar{x}^{(1)}$
and the output data $\bar{y}.$ The regularization penalty term
$E_{reg}$ is introduced to require expected properties for the
wight \cite{bio}.   The equation (\ref{loss11}) can be evaluated
by different optimizing methods such as gradient descent, Adam and
etc which in fact is an iterative method for optimization of a
function. By moving data from input layers to final layer via
feed-forward algorithm with suitable smoothness properties it
demonstrates how much predicted values are far from values of
ground truth $\bar{y}$?
    This error is then
propagated back through the network by applying back propagation
algorithm so that the weights are updated according to the amount
that they contributed to the error \cite{Rumelhart}. Predictions
are made  by providing the input to the network and by performing
a forward pass and then by generating an output.
 In
this view the architecture means how a model can be constructed
from two dimensional input data and one dimensional output
feature. With respect to the context of our problem this
architecture can be extended to more layers and neurons with
various kind of activation functions and operations of between
layers \cite{vision,speech,nlp,bio}. In the following section we
investigate correspondence between the BTZ black hole metric and
neural network components.
\section{Neural network for planar BTZ Black holes}
In 1992 Baados, Teitelboim and Zanelli investigated and obtained a
3D planer black hole which is called now BTZ black hole solution
\cite{btz}.In absence of the cosmological constant, there is no
black hole containing event horizons in 3D curved spacetimes but
thanks to the negative cosmological constant there is BTZ black
hole metric solution which provides properties similar to ones
which are appeared for 4D Schwarzschild black holes.
 By considering planar topology, general form of metric field in 1+2 dimensional black hole spacetimes is
\begin{equation}{\label{metrica1}}
ds^{2}=-f(r)dt^{2}+f(r)^{-1}dr^{2}+r^{2}dx^{2},
\end{equation}
where $x$ is a planar coordinate, $r$ is the radial coordinate and
$f(r)$ stand for the blacking functions. Einstein's field
equations can be written as
\begin{equation}{\label{eq}}
R_{ab}-\frac{1}{2}g_{ab}R-\frac{1}{L^{2}}g_{ab}=8\pi T_{ab},
\end{equation}
where $a,b=1,2,3$ in 3D spacetimes and $L$ is the AdS radius. The
right side stress tensor is assumed to be perfect non viscous
fluid such that
\begin{equation}\label{energymomentum}
T^{t}_{t}=T^{r}_{r}=-\rho, \hspace{0.3cm} T^{x}_{x}=(2w+1)\rho,
\end{equation}
where $\rho$ and $w$ are energy density and the state parameter of
the fluid respectively \cite{btz,Kiselev}. By substituting the
stress tensor (\ref{energymomentum}) and by solving the Einstein's
equations (\ref{eq}) with respect to the line element
(\ref{metrica1}) we obtain
\begin{equation}\label{solucaometrica}
ds^{2}=-\frac{r^{2}}{L^{2}}f(r)dt^{2}+\frac{L^{2}}{r^{2}}f(r)^{-1}dr^{2}+\frac{r^{2}}{L^2}dx^2,
\end{equation}
where blacking function takes on the following form.
\[
f(r)=1-\left(\frac{r_{+}}{r}\right)^{\sigma}, \hspace{0.3cm}
\sigma=2(1+w_{q}),
\]
in which $r_+$ is radius of the black hole event horizon and for
BTZ model $\sigma=2$ can be written versus the ADM mass of the
black hole $M$ and the AdS radius $L$ such that $r_{+}=\left(M L^2
\right)^{1/2}$ \cite{btz,Kiselev}. In fact the BTZ 1+2 dimensional
black hole in a Schwarzschild coordinates is stationary and
axially symmetric because it has two Killing vectors
$J^t\partial_t$ and $J^\varphi\partial_{\varphi}$
 and generically has no other symmetries for which the event horizon is determined by $M, L, J^\varphi$. In the above planner line element
 we eliminated axially symmetric property of the BTZ black hole by using the planner symmetry and so the constant angular momentum
 $J^\varphi$ is negligible.  The case $w=0$ corresponds to the non-quintessence BTZ black hole
 and $-1<w<-\frac{1}{3}$ corresponds to quintessence black hole, which in this paper we are interested for particular choices
 $w=\{0,-\frac{1}{2},-\frac{3}{4}\}$  and design artificial neural
networks in order to represent scalar field in background of them.\\
 In order to facilitate
designing neural network architecture we use the following
conformal transformation for $r$ coordinate.
\begin{equation}
dz=f^{-\frac{1}{2}}dr,
\end{equation}
in which $z$ is holographic direction and by integrating of the
above transformation we have
\begin{align} r=r_+\cosh\left(\frac{z}{L}\right).
\end{align} By substituting this into the line element (\ref{metrica1}) we obtain
\begin{align}
ds^2=-f(z) dt^2 + dz^2 + g(z)dx^2 \label{genericm}
\end{align}
where the BTZ metric components are given versus the holographic
$z$ parameter as follows.
\begin{align}{\label{metrica2}}
f(z)\equiv \frac{r_+^2}{L^2}\! \left(\sinh \frac{z}{L}\right)^{2},
\quad g(z) \equiv \frac{r_+^2}{L^2} \left(\cosh
\frac{z}{L}\right)^{2}.
\end{align}
In this conformal frame the boundary of the AdS is located at
infinity $z\to\infty$ for which $f(z)\to g(z)\approx
(r_+^2/4L^2)\exp(2z/L)\to\infty$ while the black hole horizon
lives at $z_h=0$ for which $f(z)=0$ and $g(0)=(r_+^2/L^2)$. As an
application of neural network model we like to study interaction
of a scalar field with the BTZ black hole metric as follows. We
consider a minimally coupling massive scalar field with self
interaction potential $V(\phi)$ which is propagated in the
spacetime (\ref{genericm}). Dynamics of this field is described by
the following Lagrangian density.
\begin{equation}\label{Lag}
\mathcal{L}=\sqrt{g}\bigg\{\frac{1}{2}g^{\mu\nu}\partial_\mu\phi\partial_{\nu}\phi-\frac{1}{2}m^2\phi^2-V(\phi)\bigg\},
\end{equation}
in which $g=|\det{g_{\mu\nu}}|$ is absolute value of determinant
of the metric field $g_{\mu\nu}$ and by varying with respect to
the field $\phi$ the corresponding Euler Lagrange equation reads
\begin{equation}
\square\phi+m^2\phi+\frac{\delta V}{\delta
\phi}=0,~~~\square\equiv
g^{-\frac{1}{2}}\partial_\mu(g^\frac{1}{2}g^{\mu\nu}\partial_\nu)
\end{equation}
which for (\ref{genericm}) can be written as the following first
order differential equation.
\begin{equation}
     \partial_z\pi+R(z)\pi+m^2\phi+\frac{\delta V[\phi]}{\delta\phi}=0,\label{eom}
\end{equation}
where $\pi\equiv \frac{\partial \phi(z)}{\partial z}$ is canonical
momenta of the field $\phi$ and
\begin{equation} \label{pot}R(z)=\frac{1}{2}\frac{d\ln(f(z)g(z))}{dz}=\frac{\sigma-2+2\cosh\big(\frac{2z}{L}\big)}{L\sinh\big(\frac{2z}{L}\big)}\end{equation} is an
 effective potential. This
potential is singular on the black hole horizon $z_h=0$ but has
finite value $R(\pm\infty)=\pm\frac{2}{L}$ on the AdS boundary.
The equation (\ref{eom}) together with $\pi\equiv \frac{\partial
\phi(z)}{\partial z}$ can be solved via neural network system by
discretization method. To do so the strategy should be providing a
manifestation of scalar field equation in deep neural network
scheme \cite{Has} where holographic direction $z$ mimics the deep
layers and the neurons are shown with 2 components vectors
$(\phi(z),\pi(z))$. Correspondence of the field equation with the
neural network system is possible by discretizing the equation of
motion in holographic direction  $z$ such that \cite{Has}
\begin{gather}
 \phi(z+\Delta z)=\phi(z)+\Delta z\pi(z), \notag  \\
 \pi(z+\Delta z)=\pi(z)-\Delta z\bigg(R(z)\pi(z)+m^2\phi(z)+\frac{\delta V(\phi)}{\delta\phi(z)}\bigg),
 \label{Discretization}
\end{gather}
which can be written with matrix form as follows.
\begin{equation}\label{Mat1}\left(%
\begin{array}{c}
  \phi(z+\Delta z) \\
  \pi(z+\Delta z) \\
\end{array}%
\right)=\left(%
\begin{array}{cc}
  1 & \Delta z \\
  -m^2\Delta z & 1-R(z)\Delta z \\
\end{array}%
\right)\left(%
\begin{array}{c}
  \phi(z) \\
  \pi(z) \\
\end{array}%
\right)+\left(%
\begin{array}{c}
  0 \\
  -\frac{\delta V(\phi(z))}{\delta \phi(z)}\Delta z \\
\end{array}%
\right)\end{equation}where $\Delta z$ is distance of adjacent
points in discrete coordinate system with
$z^{(n)}\equiv(N-n+1)\Delta z$, and $N$ is total number of neural
network layers. According to the figure 1 for the equations
(\ref{Mat1}) we can use $x_1\equiv\phi(z)$ and $x_2=\pi(z)$ for
components of the vector neurons $(\phi(z),\pi(z)).$ Regarding
these and linear affine transformation $x_{i}\to \sum_j
W_{ij}x_{j}$ one can obtain weights matrix $W_{ij}$ for the
equations (\ref{Mat1}) as
\begin{equation}\label{Mat2}
 W^{(n)}=\begin{pmatrix}
1&\Delta z\\
-m^2\Delta z &1-\Delta z R(z^{(n)})
\end{pmatrix}
\end{equation}
for $n$ layers and by regarding the nonlinear transformations
$x_{i}\to\varphi(x_{i})$ for each layer one can obtain activation
function for output data on each layer as follows.
\begin{equation}\label{f}
 \begin{cases}
 \varphi_1(x_1)=x_1 \\
 \varphi_2(x_2)\to x_2-\Delta z\frac{\delta V(x_1)}{\delta x_1}.
 \end{cases}
 \end{equation}
In fact the definitions (\ref{Mat2}) and (\ref{f}) bring the
scalar field system in curved geometry (\ref{Lag}) into the form
of neural network (\ref{ff}) \cite{Has}. Thus one can infer that
architect of a neural network system in this paper corresponds to
scalar field equation in
 BTZ black hole spacetime in which the weights of network play role of the BTZ black hole metric,
 $\varphi_{1,2}$
 take on role of the activation functions and holographic direction should mimic depth of the network.
 For simplicity, in the rest of the paper we set $L=1$, $m=1$ and $V[\phi]=\frac{\lambda\phi^4}{4}$
 (the Higgs potential) with $\lambda=1$ and number of hidden layers to be 8 which yields to $\Delta z=-0.1,$ $z_{b}=1$ and $z_{h}=0.1$
 (the horizon cut off frequency) which is used to regularization of interacting quantum scalar
fields. In fact input data for $\phi$ originates from quantum
fluctuations of the field (see Eq. 9 in ref. \cite{Has}) which
whose frequencies approach to infinite value on the black hole
horizon and they should be regularized. In the following section
we investigate numerical processing to produce output data or
target.
\section{The network architecture, training implementation and data setting }
\begin{figure}[!htbp]
\centering
\includegraphics[scale=0.9]{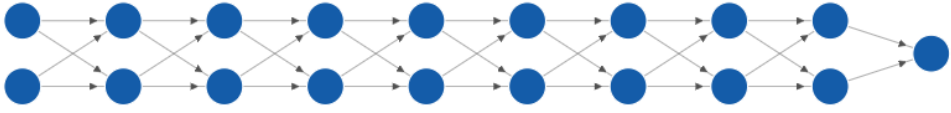}
\caption{The architectures of our neural network setup composed
with 8 hidden layers consists of two neurons in each input and
hidden layers and one neuron in output layer respectively. }
\label{q}
\end{figure}

\begin{figure}[!htbp]
\centering
\includegraphics[scale=1]{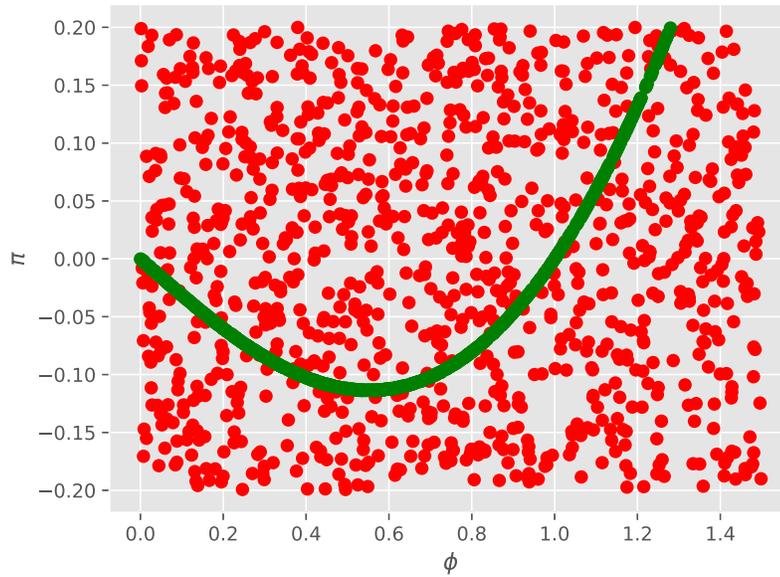}
\caption{ The data generated by the discretized BTZ metric
(\ref{pot}) to visualize how numerical values should be used as
data processing. The green points correspond to the positive data
$y_+ =0$ and the red points correspond to the negative data
$y_-=1$. This diagram is produced for non quintessence  BTZ black
hole $w=0$ and for quintessence cases $w=-\frac{1}{2},
-\frac{3}{4}$ we will have similar diagrams (not shown).}
\label{3}
\end{figure}

The architectures of our neural network setup with total 10 layers
 is shown schematically in figure \ref{q} and corresponding data
 are collected in the table 1 by designing as 8
hidden layers with two input and output layers.
The architecture is implemented via PyTorch ecosystem
\cite{Paszke} in GPU mode.
 The dataset can
be randomly produced by drawing values of independent variables
$\phi$ and $\pi$ in AdS boundary $z=1$ for domains $\phi\in[0,
1.5]$ and $\pi\in[-0.2, 0.2]$ respectively and
 transform them to the black hole horizon $z_{h}=0$ by applying the
equation of motion (\ref{Discretization}) for metric potential
(\ref{pot}) (see figure \ref{3}). To do so we choose 1000 positive
value data which can be identified by
\newcommand*\abs[1]{\lvert#1\rvert} $\abs{F}<0.1$ as cut off on the horizon
and 1000 negative value data corresponding to $\abs{F}>0.1$ which
are labeled with $y_+=0$ and $y_-=1$ respectively. In fact the
boundary condition at the horizon can be used as a classifier to
categorize generated dataset into binary classes such that for
some positive input data the output at the final layer should
satisfy
\begin{equation}
0=F\equiv\left[\frac{2}{\eta}\pi+m^2\phi+\frac{\delta
V(\phi)}{\delta\phi}\right]_{z_{fin}} \label{boundary_condition}
\end{equation}
in which $z=z_{fin}<<1$ is the horizon cutoff. Dataset will be
injected into the neural network in 200 batches. In other words we
choose 100 batches for positive and 100 batches for negative value
data respectively which they propagate through the neural network
from visible layer ($z_{h}$) to the final layer ($z_{fin}$) via
equation of motions. Our final layer is defined by the map $F$
such that the output data is $y_+=0$ for a positive answer
response data originated from quantum fluctuations of the field
\cite{Has}. In fact for limits $z_{fin}=0$ the condition
(\ref{boundary_condition})
 reads $\pi(z=0)=0.$ Now we can make the deep neural network to learn the metric
component function $h(z)$, the mass parameter of the field $m$ and
the interaction potential $V(\phi)$. The training is done by the
loss function (\ref{loss11}). In fact experiments provide only
positive answer data with $y_+=0,$ while for the training we need
also negative answer data which is to generate false response data
and so we assign output $y_-=1$ for the latter case. By according
to choice given by \cite{Has} we use a function $\tanh|F|$ for the
final layer rather than just $F$, because $\tanh|F|$ provides
$y\to1$ for any negative input. By regarding these choices the
final output of the neural network is made as binary. In this view
the activation function of final layer for cases
$w=0,~w=-\frac{1}{2}$ and $w=-\frac{3}{4}$ can respectively given
by \cite{Has},
\begin{equation}
f(F)=1+0.5\tanh[100(F-0.1)]-0.5\tanh[100(F+0.1)], \label{af1}
\end{equation}
\begin{equation}
f(F)=1+0.5\tanh[Q(F-0.1)]-0.5\tanh[0.6 (F+0.1)]. \label{af2}
\end{equation}
and
\begin{equation}
f(F)=1+0.5\tanh[1.1(F-0.1)]-0.5\tanh[P(F+0.1)]. \label{af3}
\end{equation}where $Q=\{0.6,0.9,1.1\}$ and $P=\{0.6,0.8,0.9\}.$
Looking at the figures 10 and 12 one can infer that the best fit is happened
 for $Q=0.6$ and $P=0.9$.
and to choose physically sensible metric among other
learned metrics we use lose function (\ref{loss11}) and the
penalty or regularization term given by the discrete form of the
metric potential (\ref{pot}) as
$E_{reg}=3\sum_{n=1}^{N-1}(z^i)^4(R(z^{i+1})-R(z^{i}))^2$
 to plot variation of loss function versus the leaning
rate in figure 4. This diagram shows minimum variant of the error
function is happened for learning rate $0.1$ approximately. In the
error function (\ref{loss11}) the quantities ${(\bar{x}_i,
y_i(\bar{x}))}$ are the training dataset and $\bar{y}$ is
ground-truth y. The produced errors by loss function can be saved
up across all of the training examples and the network can be
updated at the end.
The hyperparameters or training parameters which we used in this
work  are as follows: The batch size namely number of training
samples which is used to compute the gradient at each update is 10
for non-quintessence case $w=0$ and 100 for quintessence cases
$w=-\frac{1}{2}, -\frac{3}{4}$. The optimum learning rate
hyperparameter is chosen with numeric values $0.0001 $ and $0.01$
for non-quintessence and quintessence cases respectively. (These
values for learning rate can be detected by design an algorithm
which lead us to an optimal learning rate for making the model. In
the following subsection we will be talking about the procedure of
finding optimum learning rate).  Looking at the figures 5 one can
infer that presence of an suitable penalty or regularization term
is crucial to choose well learned metric among other learned
metrics. In figures 5 and 6 by using tuned values of learning rate
and batch size we have illustrated the impact of epochs on
performance of model. It can be seen in figure 6 with 50,000
epochs in which the emerged metric mimics ground-truth metric
pretty well.  To check how well the model is learned \cite{lr},
the optimization learning curves and the performance learning
curves are plotted for epoches 10,000, 30,000 and 50,000
respectively in figures 7, 8 and 9 respectively . In fact these
learning curves are as diagnostic tools for plot of model learning
optimization, performance over experience or times.Looking at
these diagrams one can infer that it is vivid the model which
learned with 50,000 epochs behaves better.
\subsection{The dynamics of learning rate}
The learning rate hyperparameter controls the speed at which the
model learns. A large learning rate allows the model to learn
faster and a small learning rate may allow the model to learn
better but the price has to be paid is longer learning time. A
learning rate that is too large could result in large weight
updates which causes the objective function of the model shows an
oscillation behavior with respect to the training epochs. The
source of oscillating behavior gets back to weights that are
diverging. On the other hand, a learning rate that is too small
may get stuck on a suboptimal solution. Diagnostic plots can be
used to investigate how the learning rate impacts learning
dynamics of the model. This is investigated by Leslie N. Smith in
\cite{lrn} in depth. He has demonstrated that if a model be
trained initially by a low learning rate and then it get increased
exponentially or linearly at each iteration a good learning rate
candidate could  be achieved but if we monitor the learning at
each iteration and then plot the logarithm of learning rate versus
loss  function, there will be spotted as the learning rate
increases and a point is appeared where the loss decreases to
stops emerges and then starts to increase again.  This minimum
point is the point we will be choosing the as the learning rate
hyperparameter of our model. In order to find minimum value of the
error function we utilized \textsc{Adam} optimizer \cite{adam}
with starting learning rate $0.1$ and corresponding exponential
decay as $\beta_1=0.05$. In fact the \textsc{Adam} optimizer is an
adaptive learning rate optimization algorithm where momentum
instead of the gradient of current step is applied to guide the
search. In other words it is combined directly as an estimate of
the first order moment of the gradient and accumulates the
gradient of the past steps to determine the direction to go. By
conducting experiment base on what explained above in order to
find optimum learning rate we obtain diagram of figure 4-a for non
quintessence case and figure 4-b for quintessence case where in
both of them a quick drop can be observed in the loss function. In
fact increasing the learning rate further will cause an increase
in the loss and even diverge from the minimum because of the
parameter updates.
\section{conclusion}
In this paper by leveraging correspondence of AdS/CFT and AdS/SL
we design deep neural network architecture for 3D
 planar BTZ and quintessence black holes to
learn boundary data which  lives on conformal field theory side.
To do so we saw that the weights of network play the role of
metric and holographic direction mimics the depth of network. Such
that data propagates from boundary to horizon of black hole and
cause to produce the background metric. We have considered a
penalty regularization term for loss function such that to be only
sensible with respect to the reality metric to be chosen among
other learned metrics. In order to achieve a high-performing
model, hyperparameters tuning has been conducted. We have noticed
loss function convergence heavily depends on the number of epochs
and learning rate. Finding faster convergence for loss function
motives us to investigate the impact of learning rate on neural
network performance by performing an experiment where we gradually
increase exponentially the learning rate to observe for steepest
drop in loss function which has guided us to pick up suitable
learning rate parameter. The message of our paper is that the
emerged spacetime could be a more universal phenomenon and helps
to understand emergence of spacetime in holographic three
dimensions. In this case one can infer that the ADS/DL
correspondence and neural network data processing paradigm could
be an applicable model instead of the unknown pure quantum gravity
theory. Such that it can say us what is happening at Planck scale
of the nature? As we saw, the error function has an integral
relationship with the emergent metric function, so the physical
parameters of the assumed black hole, such as electric charge,
angular momentum, or other physical quantities, (for instance
quintessence effect which is considered here ), will play an
important role to form the loss function and so correspondence of
two emergent metric and ground truth metric. Checking of the work
for angular momentum effect of the BTZ black hole via deep
learning and neural network data processing, is needed more time
to produce the numerical processing which we intend to do in the
next work.
 \vspace{0.5cm}
 
 \centering
\begin{tabular}{|c|c|c|c|}
\hline
Layer & Transformation & Output & dimension \\
\hline
$h_0$  & affine linear  & $(\phi_1,\phi_2)$ &  $ 2 $ \\
$h_1$  & affine linear  & $(\phi_1,\phi_2)$ &  $ 2 $ \\
$h_2$  & affine linear  & $(\phi_1,\phi_2)$ &  $ 2 $ \\
$h_3$  & affine linear & $(\phi_1,\phi_2)$  &  $ 2 $ \\
$h_4$  & affine linear & $(\phi_1,\phi_2)$  &  $ 2 $ \\
$h_5$  & affine linear & $(\phi_1,\phi_2)$  &  $ 2 $ \\
$h_6$  & affine linear & $(\phi_1,\phi_2)$  &  $ 2 $ \\
$h_7$  & affine linear & $(\phi_1,\phi_2)$ &  $ 2 $ \\
$h_8$  & affine linear & $(\phi_1,\phi_2)$ &  $ 2 $ \\
$h_{9}$  & linear  & f(F) &  $ 1 $ \\
\hline
\end{tabular}
\begin{center} Table 1. Architecture used in the networks with batch size 10.
\end{center}
\begin{figure}[ht]
\centering  \subfigure[{}]{\label{1}
\includegraphics[width=0.5\textwidth]{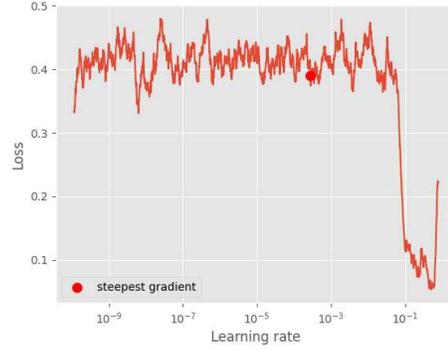}}
\hspace{3mm} \subfigure[{}]{\label{1}
\includegraphics[width=0.5\textwidth]{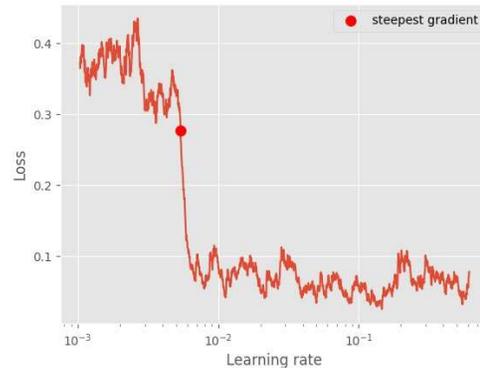}}
\hspace{3mm}
 \caption{Behavior of the loss function versus the learning rates: (a) is plotted for
non-quintessence $(w=0)$ which shows for small learning rates the
iterations become large and so minimum of the loss function is
happened at long times. (b) is plotted for quintessence
$(w=-\frac{1}{2},-\frac{3}{4})$ and it shows for large learning
rates the loss function pass far from the observed steepest
gradient point (red dot) and so the ADAM optimizer does not never
obtain minimum value for the loss function. But if we choose best
value for the learning rates equal to the steepest gradient point
the loss function reaches to its minimum value as soon.}
\end{figure}
\begin{figure}[ht]
\centering  \subfigure[{}]{\label{1}
\includegraphics[width=0.5\textwidth]{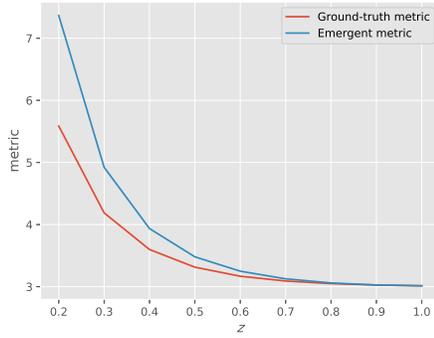}}
\hspace{3mm} \subfigure[{}]{\label{1}
\includegraphics[width=0.5\textwidth]{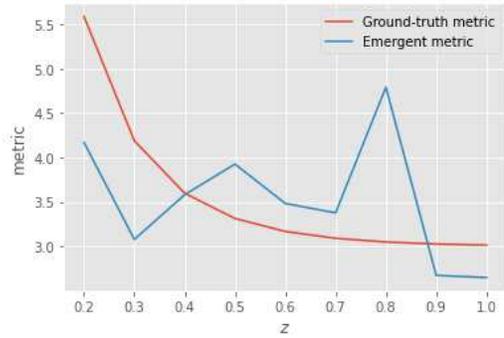}}
\hspace{3mm}
 \caption{The emerged metric and the ground-truth
metric has been portrayed with and without the penalty term
respectively at (a) and (b) after $10000$ epochs for non
quintessence case $w=0$ .}
\end{figure}
\begin{figure}[ht]
\centering  \subfigure[{}]{\label{1}
\includegraphics[width=0.5\textwidth]{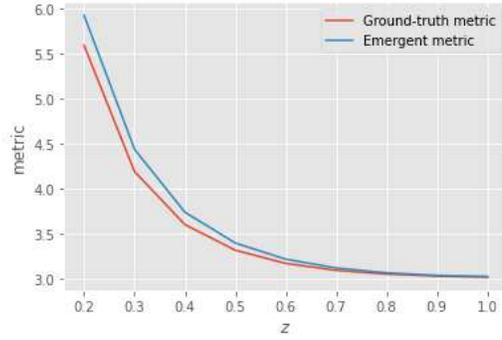}}
\hspace{3mm} \subfigure[{}]{\label{1}
\includegraphics[width=0.5\textwidth]{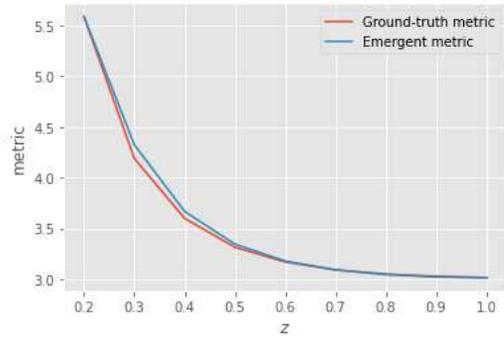}}
\hspace{3mm}
 \caption{For
case of non quintessence $w=0,$ the emerged metric and the
ground-truth metric penalty term has been portrayed at (a) and (b)
after $30000$ and $50000$ epochs respectively.}
\end{figure}
\begin{figure}[ht]
\centering  \subfigure[{}]{\label{1}
\includegraphics[width=0.5\textwidth]{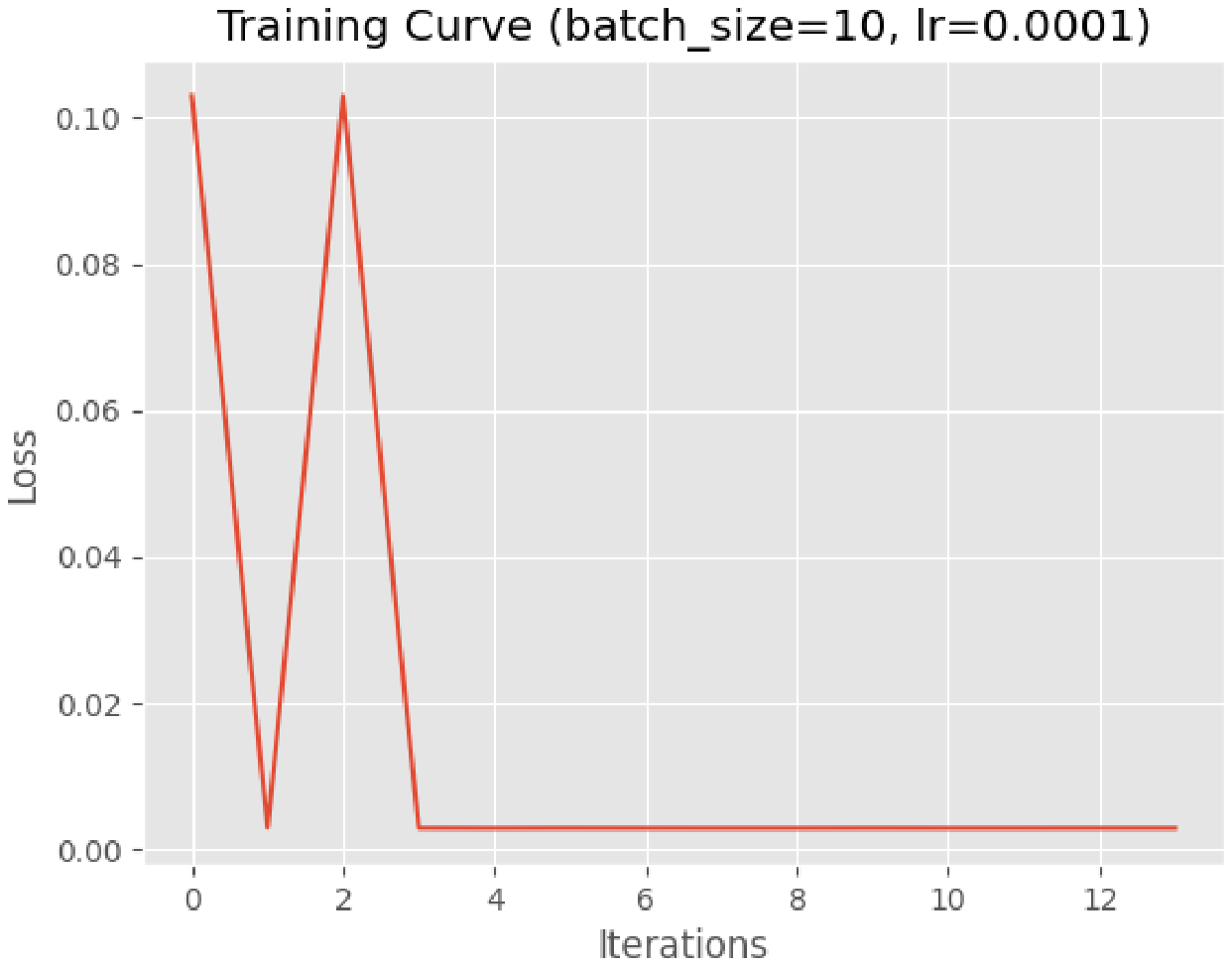}}
\hspace{3mm} \subfigure[{}]{\label{1}
\includegraphics[width=0.5\textwidth]{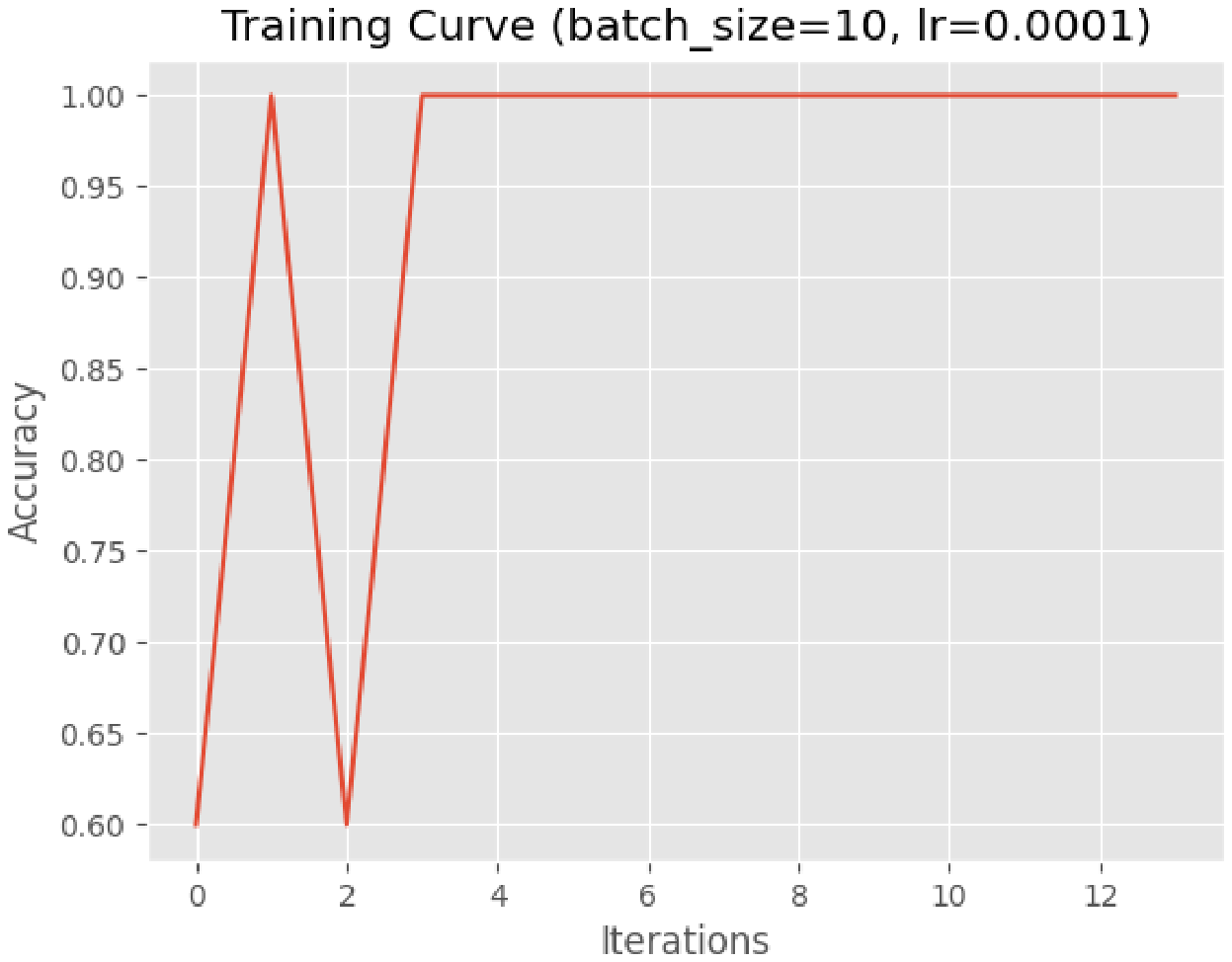}}
\hspace{3mm}
 \caption{The loss and statistical R-Squared
accuracy over $10000$ epochs. Each iteration corresponds to the
number of epochs to be over $15$. This diagrams are plotted for
non quintessence case $w=0.$}
\end{figure}
\begin{figure}[ht]
\centering  \subfigure[{}]{\label{1}
\includegraphics[width=0.5\textwidth]{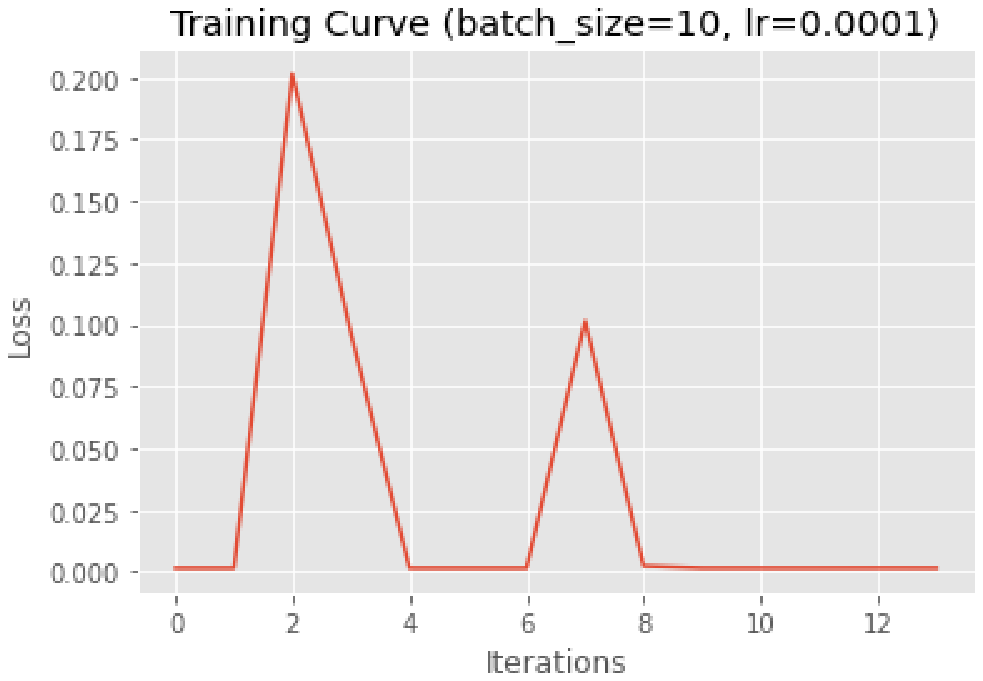}}
\hspace{3mm} \subfigure[{}]{\label{1}
\includegraphics[width=0.5\textwidth]{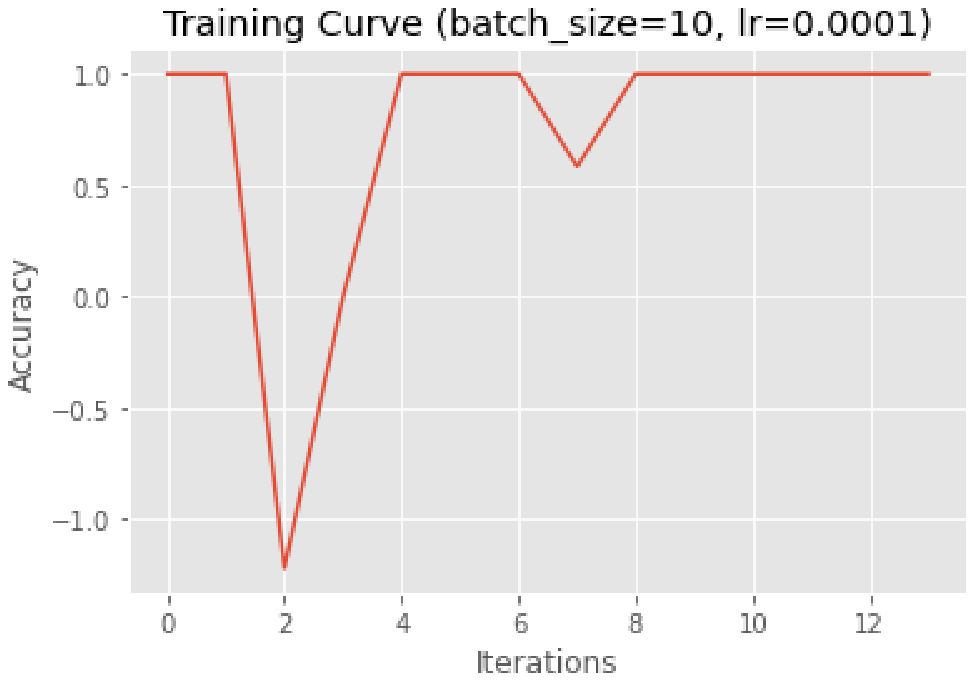}}
\hspace{3mm}
 \caption{The loss and statistical R-Squared
accuracy over $30000$ epochs. Each iteration corresponds to the
number of epochs to be over $15$. This diagrams are plotted for
non quintessence case $w=0.$}
\end{figure}
\begin{figure}[ht]
\centering  \subfigure[{}]{\label{1}
\includegraphics[width=0.5\textwidth]{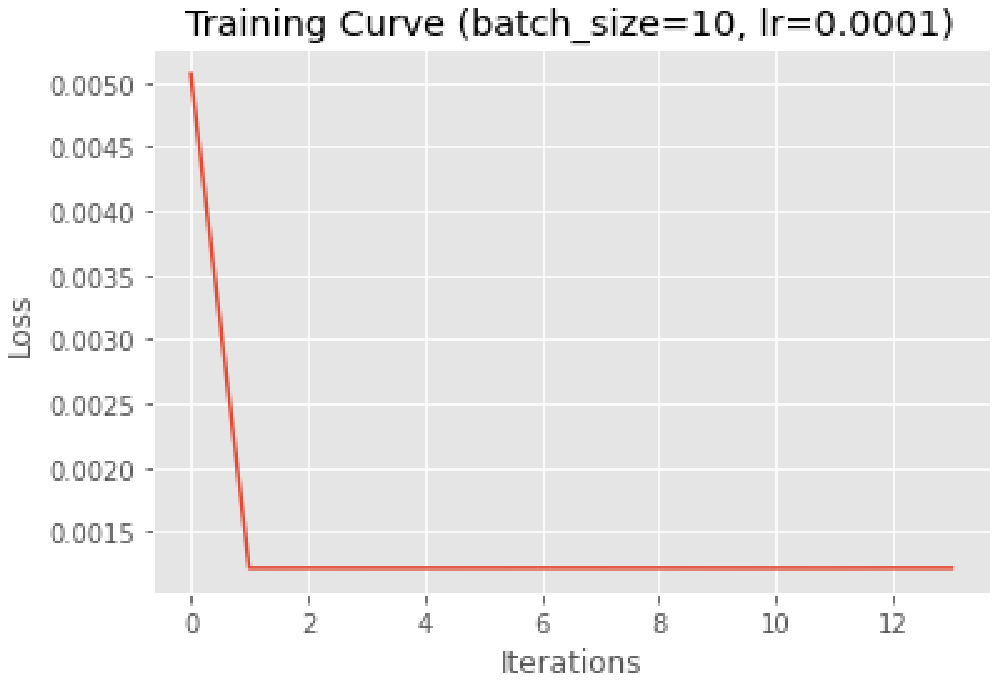}}
\hspace{3mm} \subfigure[{}]{\label{1}
\includegraphics[width=0.5\textwidth]{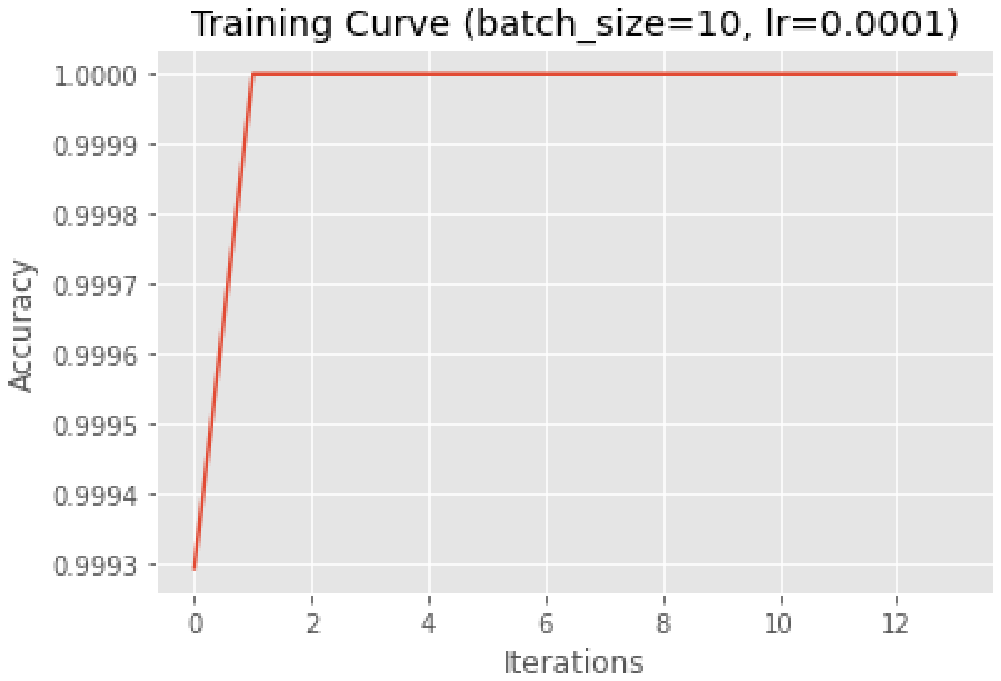}}
\hspace{3mm}
 \caption{The loss and statistical R-Squared
accuracy over $50000$ epochs. Each iteration corresponds to the
number of epochs to be over $15$. This diagrams are plotted for
non quintessence case $w=0.$}
\end{figure}
\begin{figure}[ht]
\centering  \subfigure[{}]{\label{1}
\includegraphics[width=0.45\textwidth]{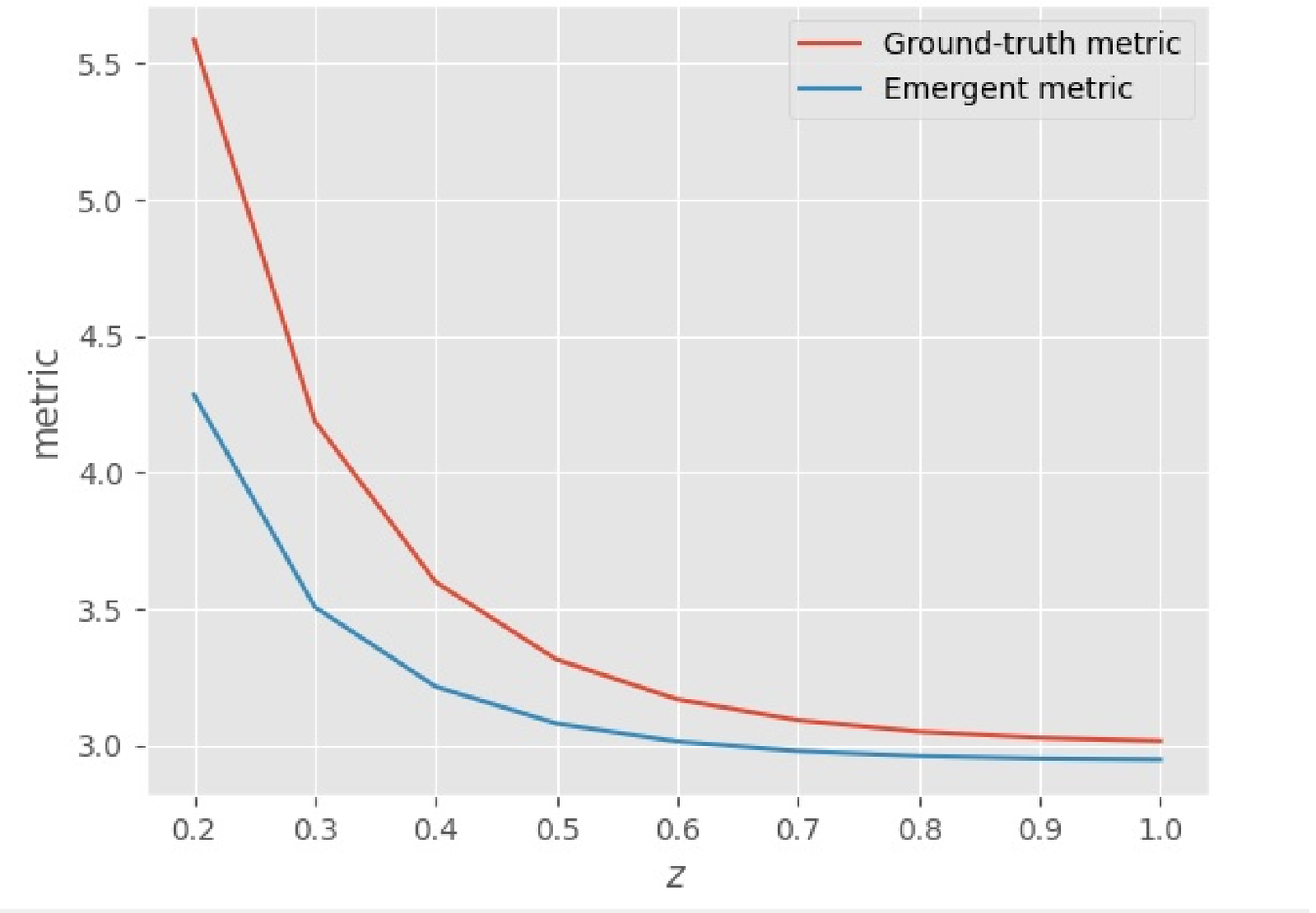}}
\hspace{3mm} \subfigure[{}]{\label{1}
\includegraphics[width=0.45\textwidth]{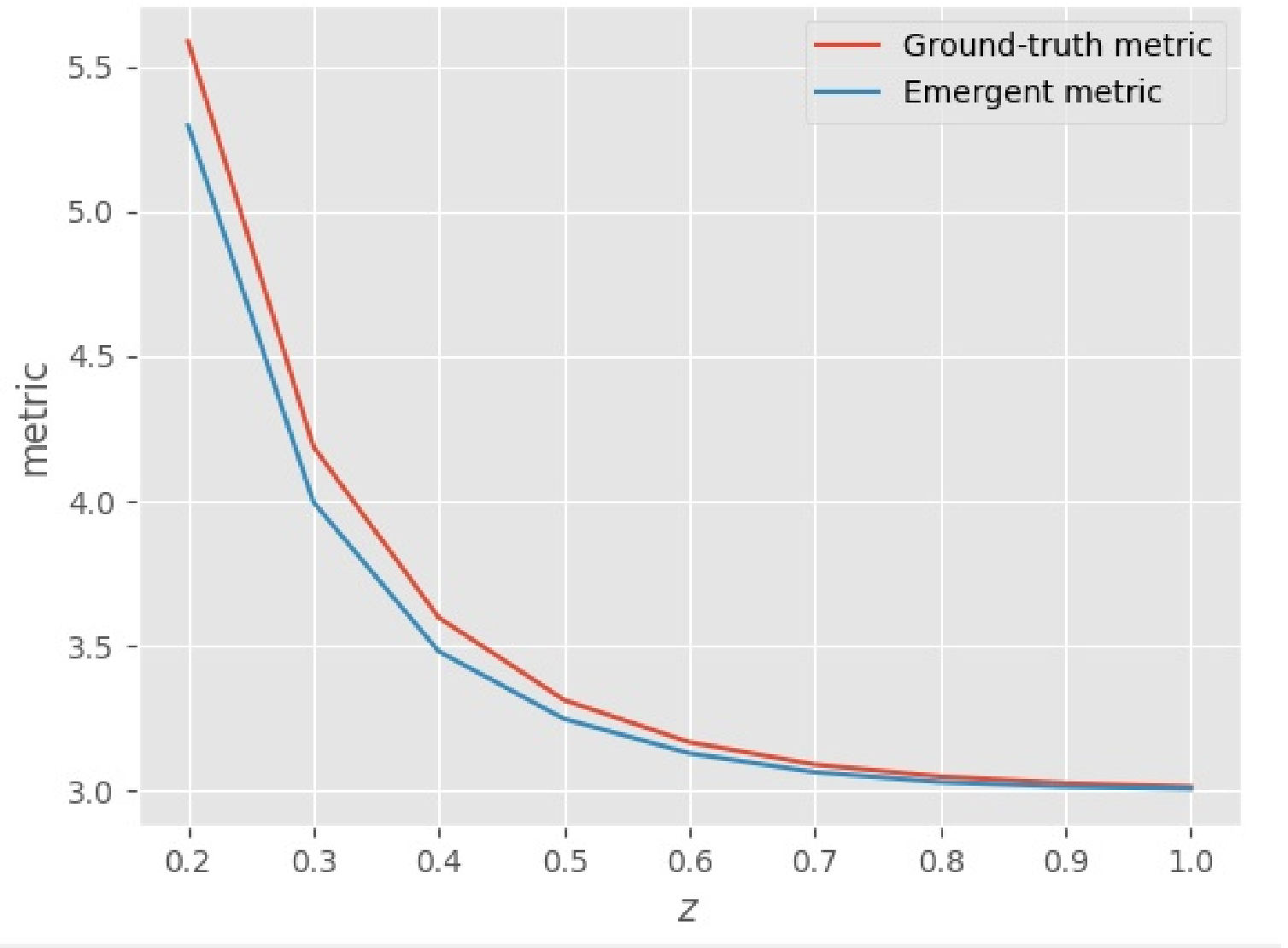}}
\hspace{3mm} \subfigure[{}]{\label{1}
\includegraphics[width=0.45\textwidth]{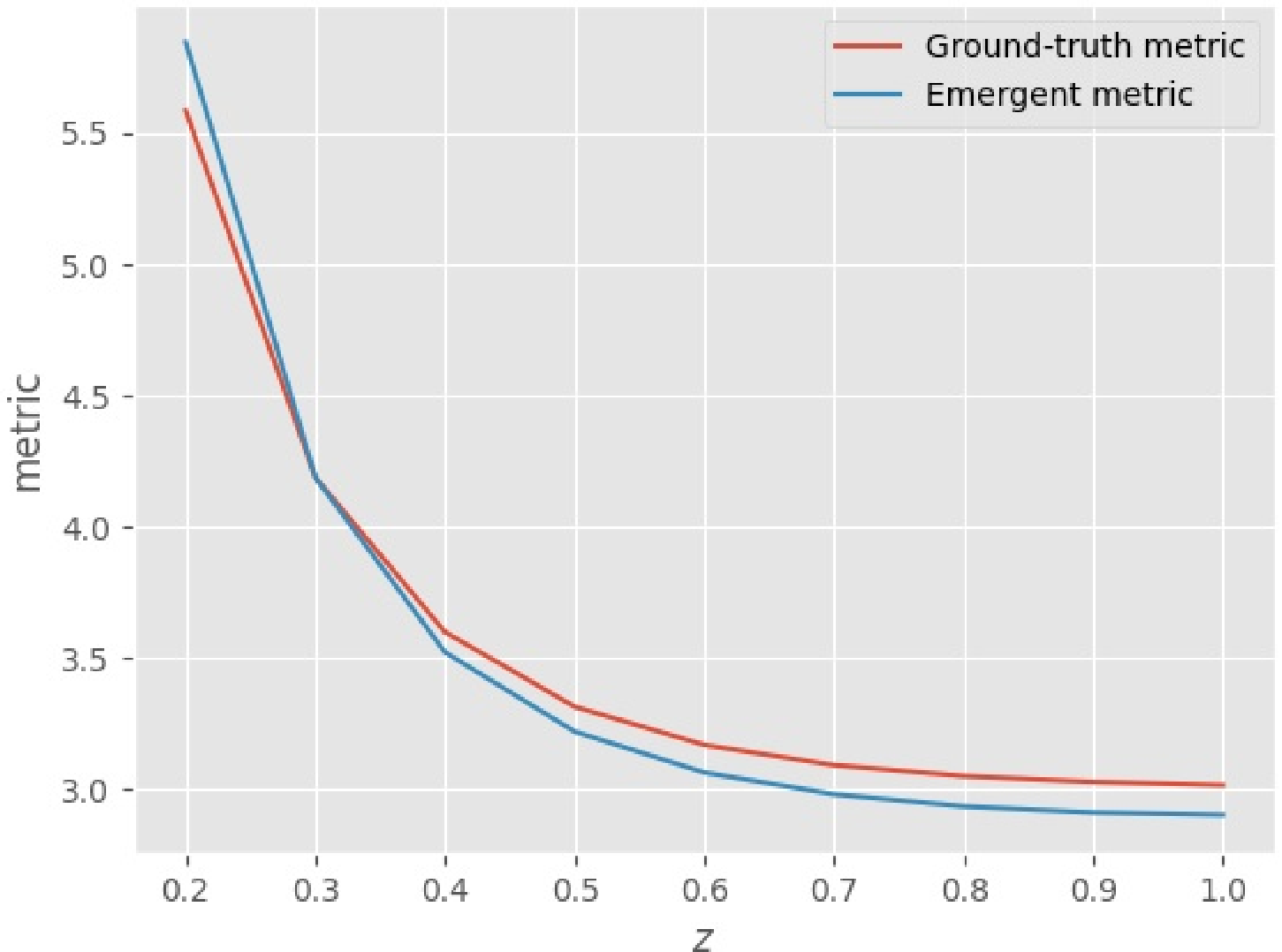}}
\hspace{3mm}
 \caption{The
learned metric with 1000 epochs with learning rate 0.01 with
quintessence case $w=-\frac{1}{2}$ for different activation
functions of last neurons given by the equation (\ref{af2}) at (a)
for $Q=0.6$, (b) for $Q=0.9$ and (c) for $Q=1.1$ respectively.}
\end{figure}
\begin{figure}[ht]
\centering  \subfigure[{}]{\label{1}
\includegraphics[width=0.45\textwidth]{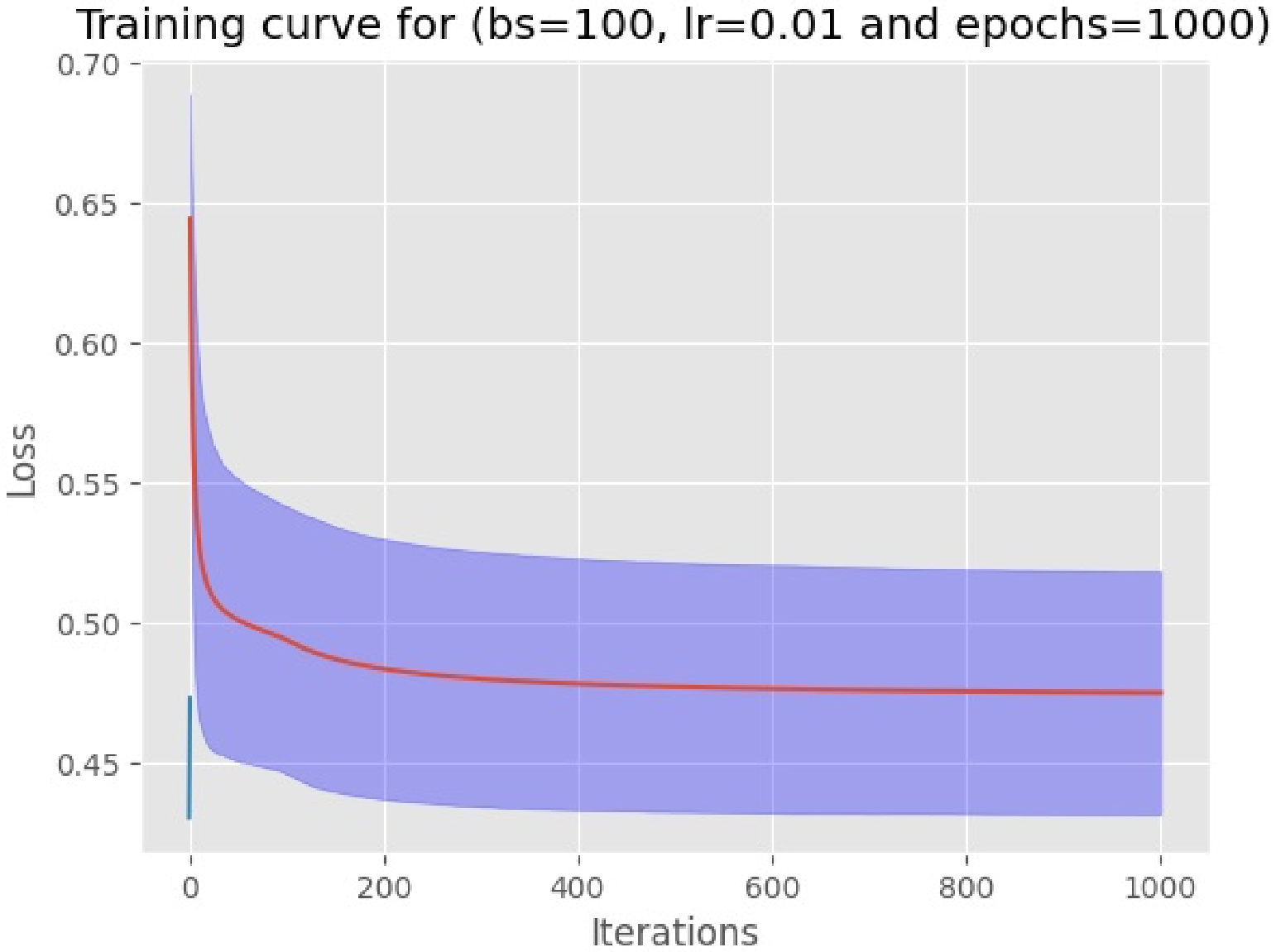}}
\hspace{3mm} \subfigure[{}]{\label{1}
\includegraphics[width=0.45\textwidth]{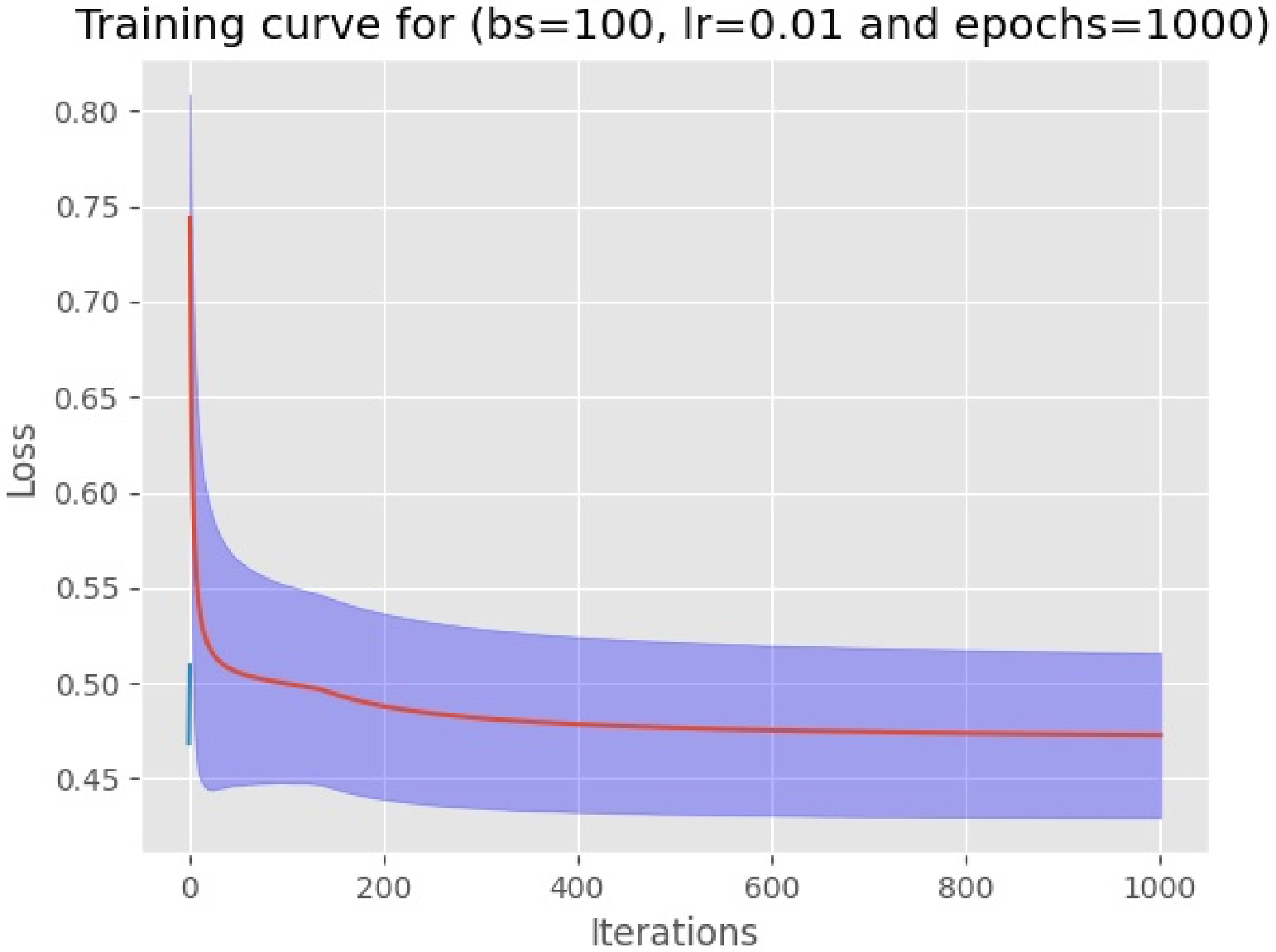}}
\hspace{3mm} \subfigure[{}]{\label{1}
\includegraphics[width=0.45\textwidth]{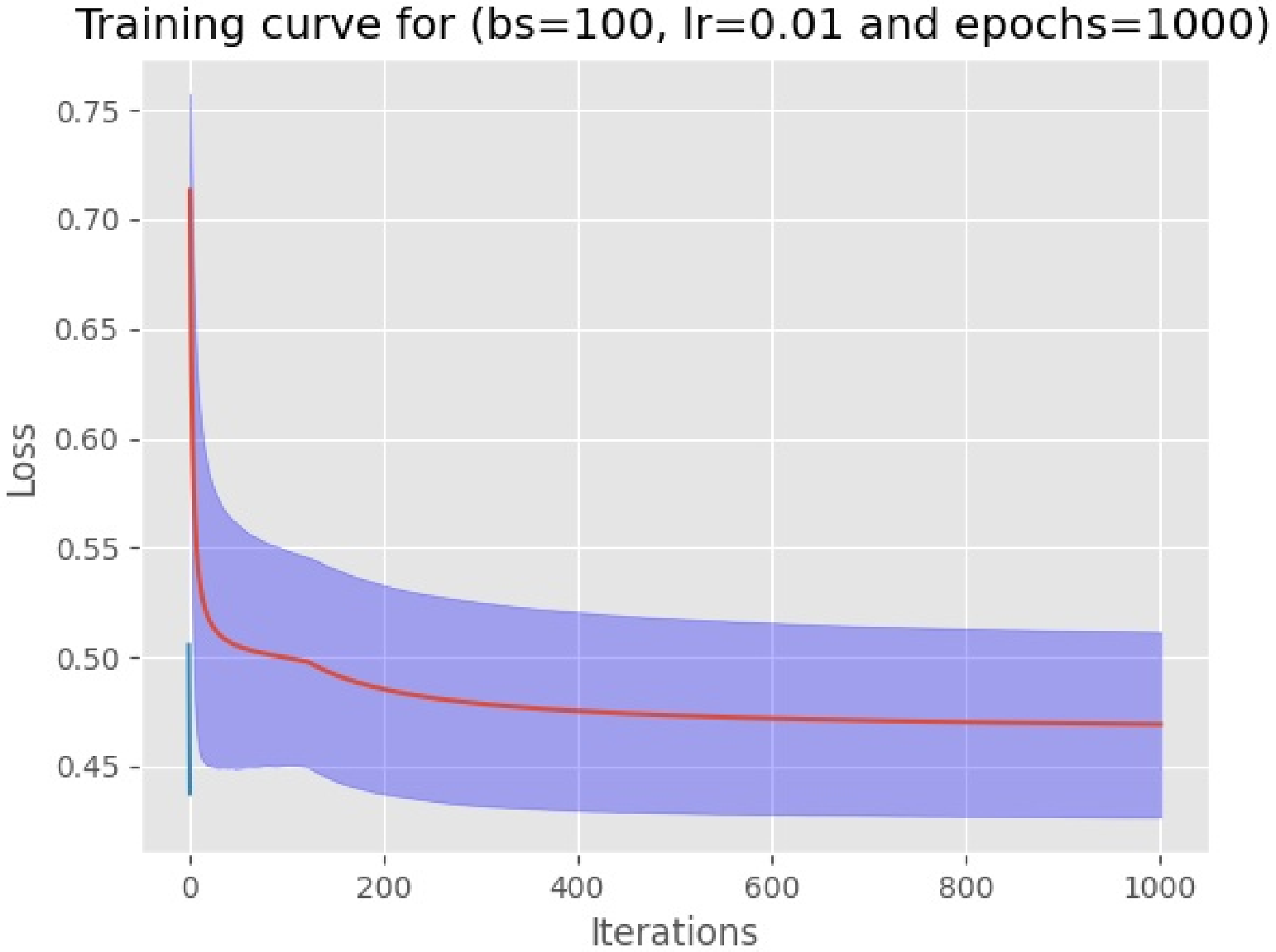}}
\hspace{3mm}
 \caption{ The behavior of the loss functions with confidence interval $95$
percent for 1000 epochs with learning rate 0.01 and quintessence
$w=-\frac{1}{2}$ for different activation functions of last
neurons given by the equation (\ref{af2}): (a) for $Q=0.6$, (b)
for $Q=0.9$ and (c) for $Q=1.1$ respectively.}
\end{figure}
\begin{figure}[ht]
\centering  \subfigure[{}]{\label{1}
\includegraphics[width=0.45\textwidth]{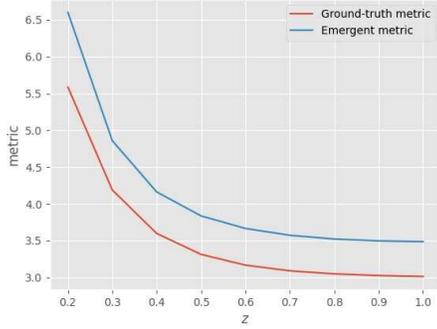}}
\hspace{3mm} \subfigure[{}]{\label{1}
\includegraphics[width=0.45\textwidth]{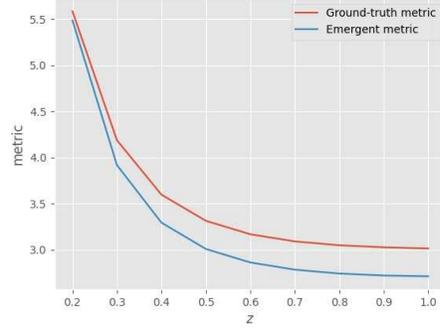}}
\hspace{3mm} \subfigure[{}]{\label{1}
\includegraphics[width=0.45\textwidth]{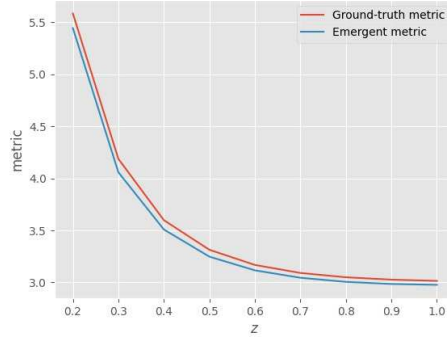}}
\hspace{3mm}
 \caption{
 The
learned metric with 1000 epochs with learning rate 0.01 for
quintessence $w=-\frac{3}{4}$ for different activation functions
of last neurons given by the equation (\ref{af3}) at (a) for
$P=0.6$, (b) for $P=0.8$ and (c) for $P=0.9$ respectively.}
\end{figure}
\begin{figure}[ht]
\centering  \subfigure[{}]{\label{1}
\includegraphics[width=0.45\textwidth]{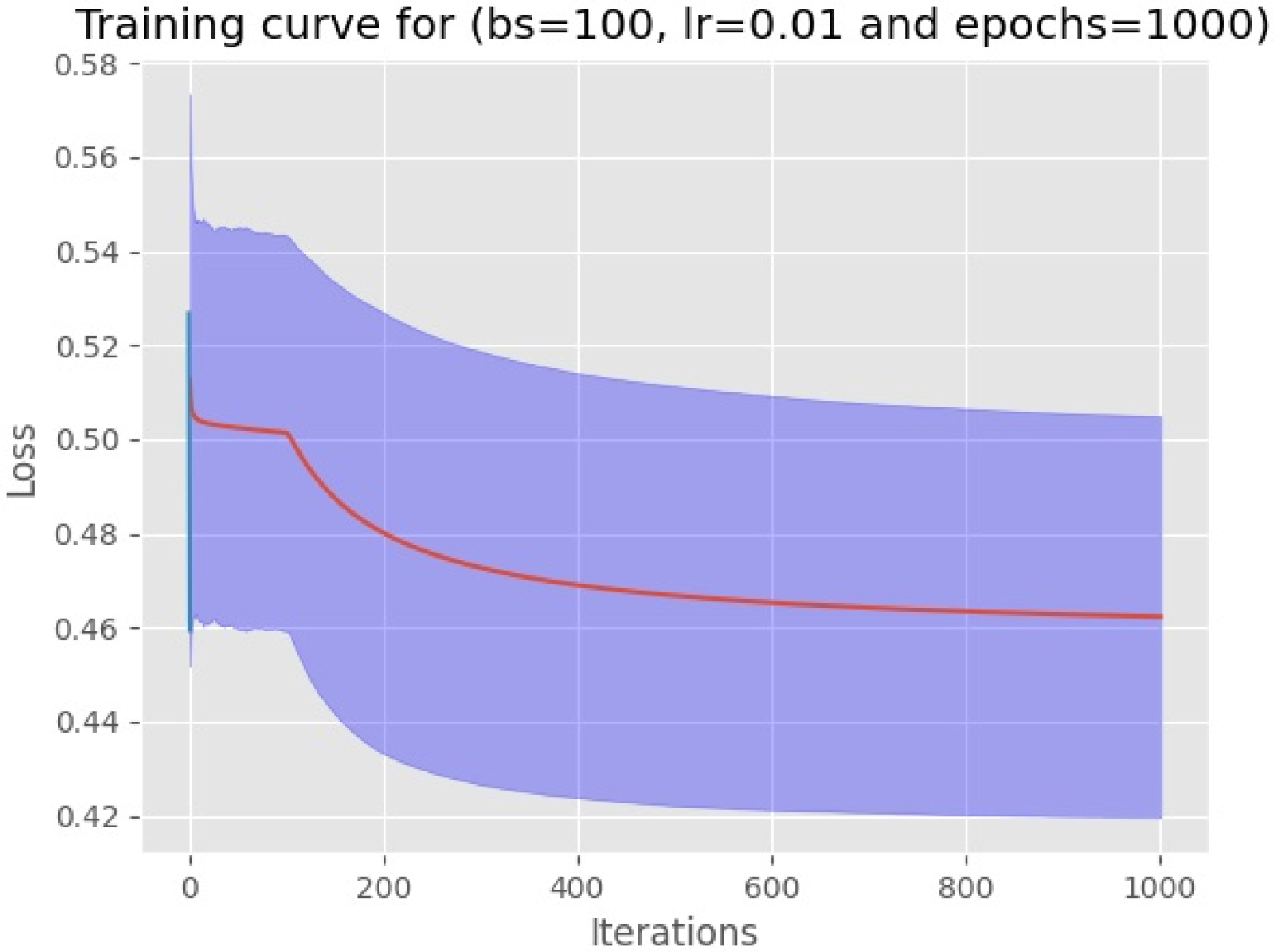}}
\hspace{3mm} \subfigure[{}]{\label{1}
\includegraphics[width=0.45\textwidth]{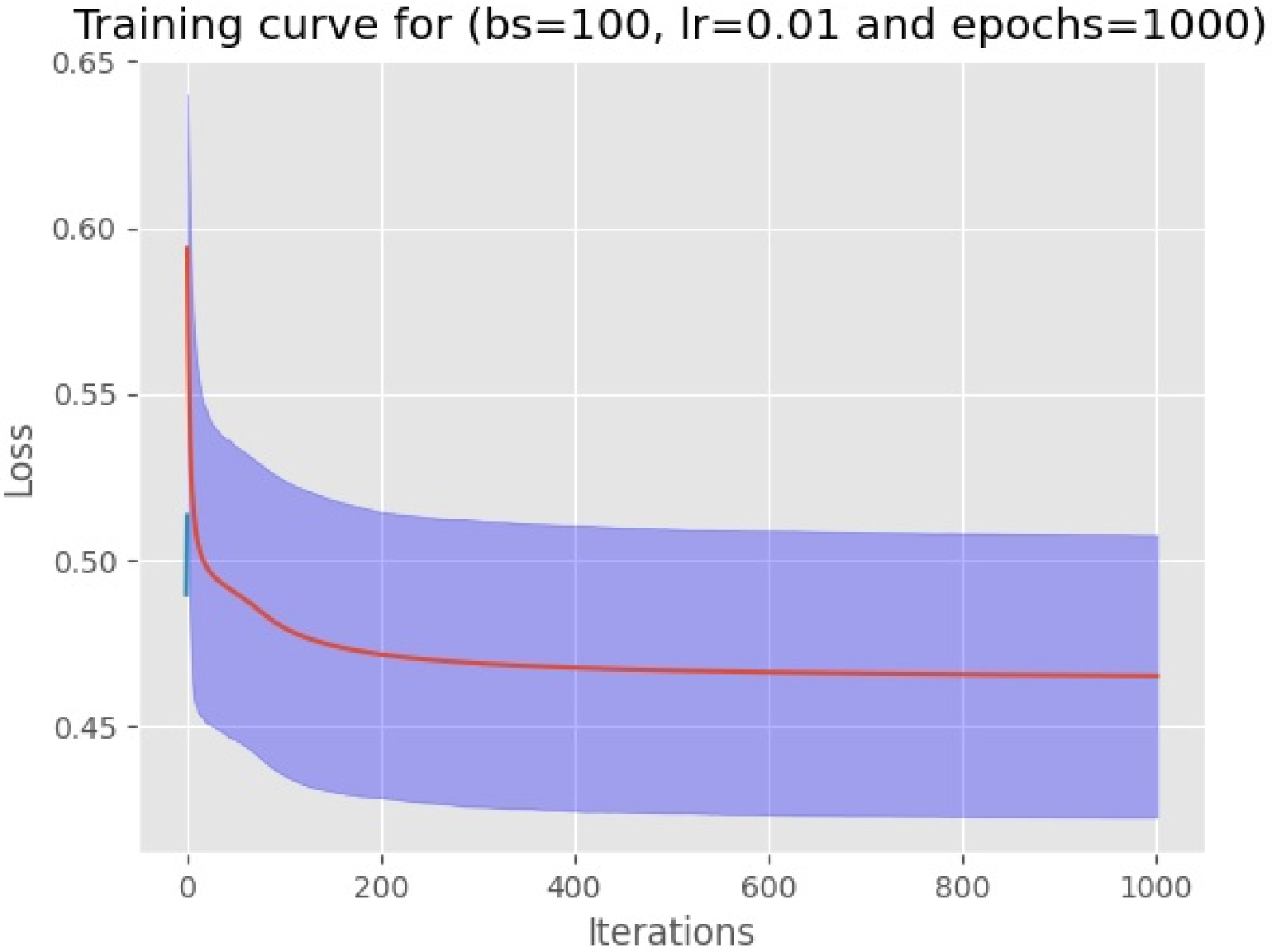}}
\hspace{3mm} \subfigure[{}]{\label{1}
\includegraphics[width=0.45\textwidth]{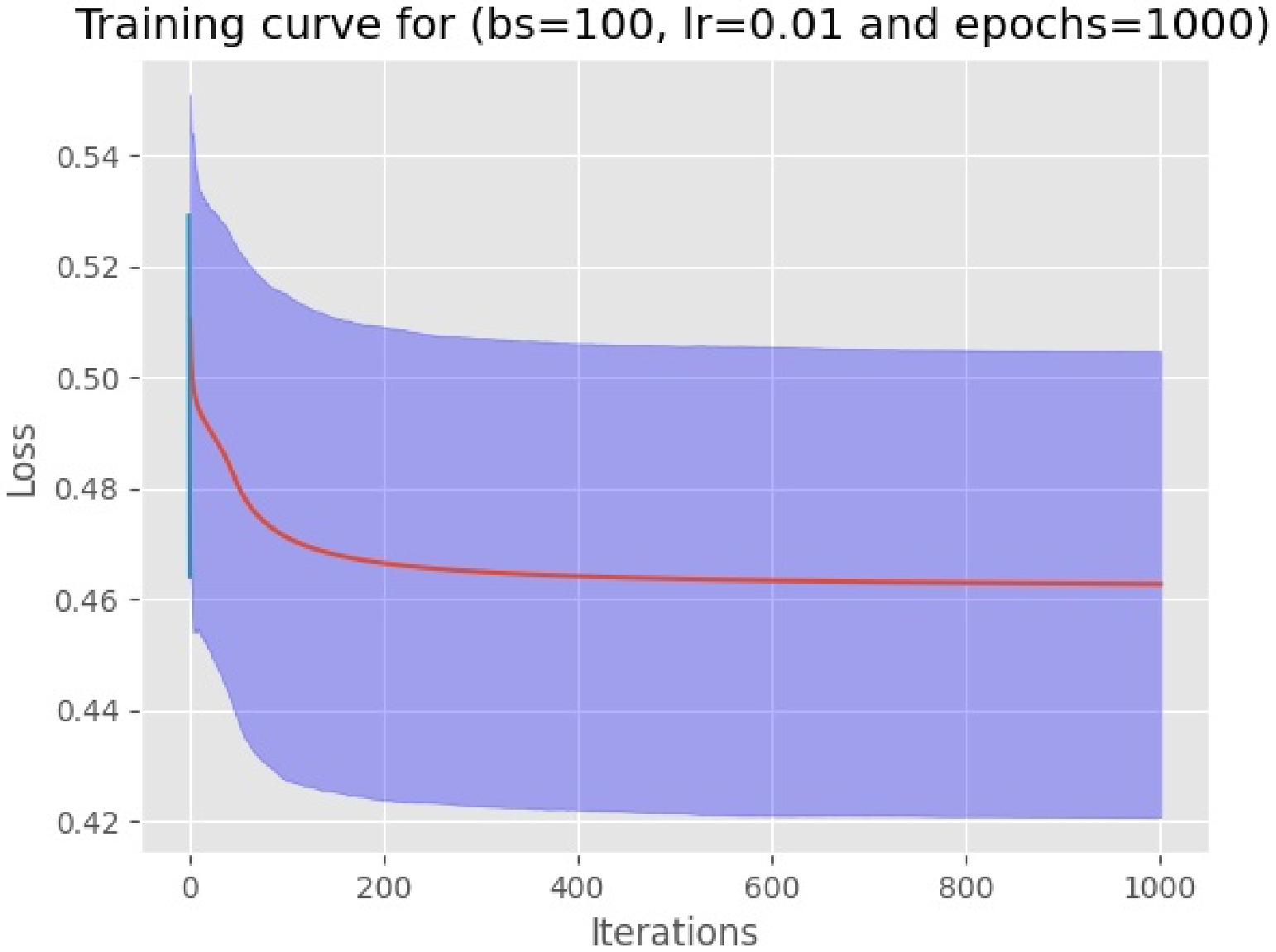}}
\hspace{3mm}
 \caption{The behavior of the loss functions with confidence interval $95$ percent for 1000 epochs
 with learning rate 0.01 and
quintessence  $w=-\frac{3}{4}$ for different activation functions
of last neurons given by the equation (\ref{af3}): (a) for
$P=0.6$, (b) for $P=0.8$ and (c) for $P=0.9$ respectively.}
\end{figure}
 
\end{document}